\documentclass[12pt,preprint]{aastex}
\shorttitle{Cosmological Aspects of GRBs}
\usepackage{epsfig}
\begin{document}

\title{Cosmological Aspects of Gamma-Ray Bursts:
 Luminosity Evolution and an Estimate of
the Star Formation Rate at High Redshifts}
\author{Nicole M. Lloyd-Ronning} 
\affil{\em Canadian Institute for Theoretical Astrophysics, \\
60 St. George St., Toronto,
Ontario, M5S 3H8, Canada}
\email{lloyd@cita.utoronto.ca}
\author{Chris L. Fryer}
\affil{\em Theoretical Astrophysics, Los Alamos National Labratories, \\
Los Alamos, NM 87544}
\author{Enrico Ramirez-Ruiz}
\affil{\em Institute of Astronomy, University of Cambridge,\\ Madingly Road, Cambridge,
CB3 0HA, U.K.}

\begin{abstract}

Using 220 Gamma-Ray Burst (GRB)
redshifts and luminosities  
derived from the luminosity-variability relationship of Fenimore \&
Ramirez-Ruiz (2000), we show that {\em there exists a significant
correlation between the GRB luminosity and redshift.}  In particular,
we find that the evolution of the average luminosity  
can be parameterized as 
$L \propto (1+z)^{1.4 \pm \sim 0.5}$, where $z$ is the burst redshift.  We
discuss the possible reasons behind this evolution and compare it to
other known sources that exhibit similar behavior.  In addition,
we use non-parametric statistical techniques to independently estimate
the distributions of the luminosity and redshift of bursts, accounting
for the  evolution (in contrast to previous studies which
have assumed that the luminosity function is independent of redshift).
We present these distributions and discuss their implications.  Most
significantly, we find a co-moving rate density of GRBs that continues
to increase to $(1+z) \ga 10$. From this estimate of the GRB rate
density, we then use the population synthesis codes of Fryer et al. (1999)
to estimate the star formation rate at high redshifts, based on
different progenitor models of GRBs.  We find that no matter what the
progenitor or population synthesis model, {\em the star formation rate
increases or remains constant to very high redshifts} ($z \ga 10$).
  
\end{abstract}
\section{Introduction}
A luminosity function and redshift distribution are among the most
sought after quantities for any class of astrophysical objects.  From
QSOs, to various types of galaxies, to supernovae, etc., these
distributions provide important insights not only into the physics of
the individual objects themselves, but also into the evolution of
matter in our universe.  A luminosity function is a measure of the
number of objects per unit luminosity and therefore is intimately
connected to the energy budget (e.g. mass, rotational energy, etc.)
and the physical parameters determining the emission mechanism
(e.g. density, magnetic field, etc.) of the objects.  On the other
hand, the co-moving rate density, a measure of the number of events
occurring per unit co-moving volume and time, provides a census of the
number of objects formed at a given redshift and can help us
understand object/structure formation in its various stages of
evolution.  Often, when doing large statistical studies of a
particular class of objects, the luminosity function and redshift
distribution are assumed to be independent quantities; that is, the
sources' luminosity function is assumed to be the same for all
redshifts.  This makes the analysis easier when one has limited
information (e.g. significant data selection effects) obscuring a
direct interpretation of the measured distributions of luminosity and
redshift.  However, it has been shown that this assumption is not
valid for many astrophysical objects (for example, see \S 5.1.3)
 
In this paper, we examine the evolution of the luminosity
function, as well as the co-moving rate density of Gamma-Ray Bursts.
Currently, there are on the order of about $ 20$ bursts with measured
redshifts (see Jochen Greiner's homepage at:
http://www.aip.de/$\sim$jcg/grb.html for a compilation of all observed
afterglow results); a histogram of all the redshifts as of July,
2001 is shown in Figure 1. 
\begin{figure}[t]
\centerline{\epsfig{file=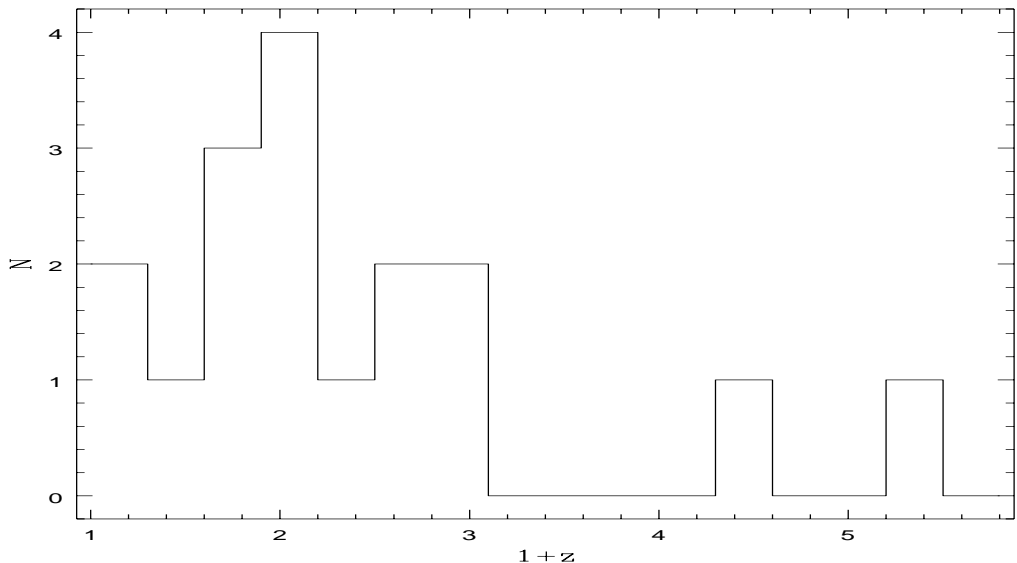,width=0.9\textwidth,height=0.49\textwidth,angle=0}}
\caption{Histogram of seventeen bursts with confirmed measured redshifts.}
\end{figure}
 The gamma-ray burst community has been very 
successful in gaining a working understanding of GRBs from these (and 
a few other) individual bursts.  By studying the prompt and especially 
afterglow emission in great detail for a few GRBs, some important 
generalizations have been made.  For example, it was clear from 
the very first bursts with measured redshifts that the GRB
luminosity function is {\em not} very narrow and in fact exhibits a
rather large dispersion (unless otherwise stated, by ``luminosity'',
we mean luminosity modulo a beaming factor $d\Omega$). Through
indirect evidence (extinction estimates, position in host galaxy,
possible SNe in light curves; see \S 5 for further discussion), it can
be inferred that many GRBs are associated with star forming regions
which helps constrain certain progenitor models (see, e.g., Meszaros,
2001, and references therein).
 
However, there are still a number of unanswered questions that can
only be addressed through larger scale population studies of GRB
luminosities and redshifts.  Because GRBs are most likely associated
with some sort of compact object(s) (e.g. a very massive star or two
compact objects such as neutron stars or black holes merging), the
luminosity distribution may help us eventually constrain the
parameters on which the energy output depends (such as mass and
angular momentum), and ideally help us understand the nature of the
progenitor.  If this luminosity function evolves, this not only
provides additional insight into the physical parameters fundamental
to the gamma-ray burst, but may help shed light on the evolution of
the GRB host environment.  Finally, because of a probable connection
to massive stars (in {\em both} the collapse and merger models), the
GRB co-moving rate density can provide novel information about the
rate of star formation in our universe.

Therefore, getting a handle on the GRB luminosity function and rate
density could have important consequences for understanding not only
GRBs themselves, but other astrophysical problems as well
(particularly, determining the star formation rate at high redshifts).
But because the GRB luminosity function is not a
$\delta$-function, we cannot use its flux as a standard candle from
which to infer a redshift to the source (for example, see Figure 1 in
Lloyd et al., 2000, which shows the dispersion in the isotropic
equivalent energy for those bursts with measured redshifts).  At first
glance, it appears that direct measurements of a large number of
redshifts to bursts (e.g. by HETE-2 and Swift) with careful accounting
of the detector selection effects is the only way to get a handle on
the burst density distribution and luminosity function.
 
However, recently it has been suggested that there exist other types
of standard candle or Hubble-type relationships in GRBs, from which a
redshift can be inferred from a common GRB observable.  For example,
Norris, Marani, \& Bonnell (2000) report a positive correlation
between spectral lag and luminosity - from 6 bursts with measured
redshifts, the higher luminosity bursts show more of a delay between
the high energy and low energy pulses than do weaker bursts. Fenimore
\& Ramirez-Ruiz (2000; hereafter, FRR) were the first to report a
relation between the bursts variability, $V$, - defined by the
"spikiness" of the light curve - and the luminosity for $8$ bursts
with measured redshifts (see also Reichart et al., 2000, for further
analysis of this relation).  From this
relationship, they estimated luminosities and redshifts to 220
GRBs with a peak flux above $1.5 \rm ph/cm^{2}/s$.  This was done by
computing the variability (see FRR for exact definition) of each
bursts' light curve, then inferring the luminosity from the determined
correlation, and finally computing a redshift (assuming some
cosmological model).  Their procedure is described in further detail
below.
 
{\em The goal of this paper is to test for correlation between
luminosity and redshift in the FRR data, estimate the bivariate GRB
distribution of luminosity and redshift $\Psi(L,z)$, and then use
this distribution to place constraints on the star formation rate at
early times.}  It is important to point out that because of the flux
threshold in the FRR sample, there is a strong truncation of the data
produced in the $L-z$ plane (see \S 2.2 and 3), which renders it difficult
to estimate the underlying parent (or intrinsic) distributions of $L$
and $z$, or any intrinsic correlation between them.  FRR attempt to
account for this truncation threshold by first assuming {\em
independence} of the variables $L$ and $z$, and then constructing
redshift-luminosity bins parallel to the $L$ and $z$ axes. They
estimate a luminosity function $\phi(L)$ in each redshift bin, and
then find a function $\rho(z)$ that causes the $\phi(L)$ curves in
each redshift bin to fall on top of one another or match each other.
Using this method, FRR find $\phi(L) \propto L^{\sim -2.3}$ and
$\rho(z) \propto (1+z)^{3.3 \pm 0.3}$.   
We emphasize
that their method {\em implicity assumes no luminosity evolution}\footnote{We 
do point out that when {\em assuming} a density distribution
proportional to the star formation rate, FRR do find the luminosity
function must evolve, although do not quantify this.}.
 
In this paper, we show that if the $L-V$ luminosity indicator
is valid, it leads to a GRB luminosity function that depends on
redshift.
 We use robust non-parametric statistical
techniques to estimate the significance of this correlation given the
selection threshold.  Accounting for this luminosity evolution, we
then investigate the density distribution and
shape of the luminosity function for this sample, and use this
information to make an estimate of the star formation rate
at high redshifts.  The organization
of this paper is as follows: In \S 2, we give a brief introduction
into the nature of the luminosity-variability correlation, and discuss
the data sample we use in our study.  In \S 3, we describe the
statistical techniques employed to account for the flux threshold
truncation and then apply these to our sample.  In \S 4, we describe
our results. We find that the burst luminosity $L$ is significantly
correlated with redshift $z$ and can be parameterized by $L \propto
(1+z)^{\sim 1.4}$.  Accounting for this correlation, we estimate
the redshift distribution as well as the distribution of a measure of luminosity
(with redshift dependence removed).  In particular, we find that the
co-moving rate density of GRBs increases to very high redshifts.  In
\S 5, we discuss the implications of GRB luminosity evolution and
meaning of the luminosity and density distributions.  We then use
population synthesis codes developed by Fryer et al.(1999) to estimate
the star formation rate at high redhift from the GRB co-moving rate
density.  We find that the star formation rate increases at high
redshift for merger models, and either increases or remains
approximately flat at high redshift for collapsar models.  Therefore,
no matter what the GRB progenitor we use, {we find the star
formation rate continues to increase (or at the very least remains
constant) beyond a redshift of $(1+z) \sim 3$}.  In \S 6, we summarize
and present our main conclusions.

\section{The Luminosity-Variability (L-V) Correlation}
 
As described in the introduction, Fenimore \& Ramirez-Ruiz (2000) have
found a correlation between the GRB luminosity (in the 50-300 keV
energy range\footnote{Using the ``bolometric''
gamma-ray luminosity gives similar results, however.}), 
and the light curve ``variability'' for 8 bursts with
measured redshifts (models for the physical mechanism
behind this correlation can be found in Plaga, 2001, Ioka \& Nakamura, 2001,
Kobayashi et al., 2001, and Ramirez-Ruiz \& Lloyd-Ronning, 2002).
This relationship is confirmed by Reichart et al. (2000) using
a slightly different measure of variability as well as other subtle
differences as described in \S 3 of FRR.  In their most recent
analysis, FRR find that the luminosity $L$ is proportional to a
measure of variability $V$ of the GRB lightcurve (a further
description of variability is given in \S 2.2 below; for the exact
definition, see equation 2 of FRR) to a power $1.57 \pm \sim 0.5$.
\footnote{Earlier versions of the paper which used a different degree of
smoothing of the light curve, and therefore a different
definition of the variablity $V_{*}$ gave $L \propto V_{*}^{3.35 (+ 2.45,
-1.15)}$.  It is important to point out that no matter which 
luminosity indicator we use (with variability defined as $V$ or
$V_{*}$), all of our results
remain qualitatively similiar.}  Because $V$ is a readily measured
quantity from the bursts' light curves, the $L-V$ relation serves
as a luminosity indicator for those bursts which {\em do not}
have measured redshifts.  Once the luminosity is found from this
relationship, a redshift can then be computed (under the assumption
of some cosmological model).

As seen in the papers of FRR and
Reichart et al. (2000), the correlation has significant scatter.  However,
although the exact power-law relationship is not well established, its
existence appears to be robust. For example, Schaefer (2001) finds a
relationship between pulse lag and variability in the BATSE data which
is predicted if both the lag-luminosity and variability-luminosity
correlations are real.  In addition, several authors have found
evidence that appears to support this relationship in the GRB light
curve power density spectrum (PDS).  Pozanenko (private comm.)  finds
a correlation between the GRB luminosity and ``RMS variability'', where
the RMS variability is the integral of the PDS above some fixed
frequency (and therefore a measure of the contribution of high
frequency emission to the lightcurve).  Similarly, Beloborodov et
al. (2000) find a positive correlation between the PDS spectral index
and luminosity for bursts with redshifts.  The spectrum is shallower
for brighter bursts (implying more high frequency emission for
brighter bursts), which is also at least qualitatively consistent with
the existence of the luminosity-variability correlation. As mentioned
above and as we show below, despite the scatter in the FRR
luminosity-variability relation, the qualitative nature of our results
are not sensitive to the precise value of the $L-V$ correlation index.
In this section we describe the data sample obtained from this
correlation, which we use in our investigation.

\subsection{Data} 
The data we use in this analysis is from Table 2 of FRR, which lists the peak flux in
the 256ms time bin and from 50-300keV, the $T_{90}$ duration, the
measure of the variability, the determined redshift and luminosity from 
the empirically determined relationship, for 220 BATSE bursts.  From 
the burst time profile, they calculate a measure of variability,
defined in their equation (2).  The variability is most easily
understood as the degree of "spikiness" of the light curve - that is,
the sum of mean square difference between the actual counts and a
smoothed light curve. Their definition of variability accounts for
time dilation and the fact that the time scale in a GRB light curve is
energy dependent (Fenimore et al., 1995) as well as for the presence
of Poisson noise (the expected value of $V$ for pure Poisson noise is
zero).  Then, given the relationship between luminosity and
variability presented in their paper,
$L/d\Omega=3.1\times10^{56}V^{1.57}$ erg s$^{-1}$, they calculate a
luminosity for the 220 BATSE GRBs in their sample. Finally they solve
for redshift through the relation \begin{equation} L/d\Omega = f_{256}
<E> [\int_{0}^{z}(c/H_{o})\frac{dz}{\sqrt{\Omega_{\Lambda} +
\Omega_{m}(1+z)^{3}}}]^{2}
\end{equation}
where $f_{256}$ is the peak photon flux on the 256 ms timescale and in
the energy range $50-300$ keV, $<E>$ is the average energy of the
spectrum obtained by integrating the spectrum over the $50-300$ keV
range, $c$ is the speed of light, and they use $H_{o}=65 \rm km/s/Mpc$, $\Omega_{m}=0.3$,
$\Omega_{\Lambda}=0.7$.\footnote{Note the factors of $(1+z)^{2}$ present in
the usual definition of luminosity distance and explicitly absent in
the expression above are contained in the $<E>$ calculation; see FRR
for a discussion of this luminosity distance issue.}  Because of the
subtleties of the dependence of the light curve on redshift (through
time dilation and the dependence of pulse width on energy), the
redshift is actually obtained through an iterative procedure as
described in \S 4 of FRR. Note that they choose an upper limit to the
redshift of $12$.
 
\begin{figure}[t]
\centerline{\epsfig{file=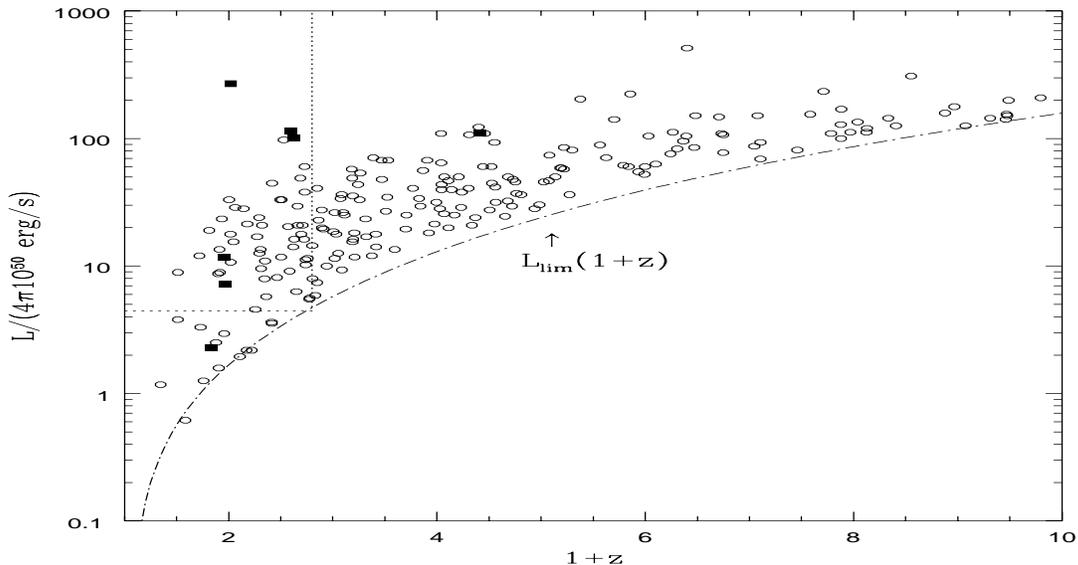,width=0.9\textwidth,height=0.49\textwidth,angle=0}}
\caption{Normalized luminosity vs.redshift for 7 bursts (squares) with confirmed redshift measurements
from which the L-V correlation of FRR was derived, and the 220 BATSE bursts (circles)
on which
this correlation was applied to obtain luminosities and redshifts (see \S 2 of text).
The dot-dashed line is the truncation limit as a result of the peak flux
threshold in the FRR sample. The dotted vertical and horizontal lines show a representative
``associated set'' for one particular burst
as used in our statistical methods described in \S 3.}
\end{figure}
The FRR sample is chosen to have peak fluxes above the threshold
$f_{256}=1.5 \rm ph/cm^{2}/s$, which defines a limiting luminosity as a
function of redshift according to the equation above.  The data
(circles) and their truncation limit as defined by this equation are
shown in Figure 2, where the solid squares denote those bursts with
measured redshifts; the horizontal and verticle lines are relevant for
our statistical methods and are explained in \S 3.  The truncation
places an important restriction on the amount of information we can
obtain from the data.  Because our goal is to learn about the {\em
parent} (or intrinsic) distributions of $L$ and $z$ (e.g. the
distributions without truncation), we must account for the trunction
limit in some way.  Fortunately, because the threshold is so well
defined, we can use previously developed non-parametric techniques to
gain full information on the underlying distributions from the
observed distributions and knowledge of the truncation limit.

\section{Statistical Techniques}
Our goal is to estimate the bivariate distribution of luminosity and
redshift, $\Psi(L,Z)$, where $Z=1+z$.  We do not assume $L$ and $Z$ are
independent.  However, without loss of generality, we can write
$\Psi(L,Z) = \rho(Z)\phi(L/\lambda(Z))/\lambda(Z)$, where $\phi(L/\lambda(Z))$
is the luminosity function, and $\lambda(Z)$
parameterizes the correlation between $L$ and $1+z$.  Trying to
estimate {\em correlations between} and {\em distributions of}
variables which suffer from selection effects without accounting for
the truncation can lead to drastically false conclusions.  Obviously,
if one blindly computed the correlation between $L$ and $Z$ without
accounting for the severe truncation present in the $L-Z$ plane due to
the lower flux threshold as seen in Figure 2, one would find they are
highly correlated.  Of course, this would be a result of the
truncation itself, and not reflect any real physics in the underlying
distributions.  The point is that we need some way of learning about
the data not observed (falling below the truncation line) from the
data that is observed and from the truncation itself.
  
There are simple, straightforward ways to estimate the parent
distribution of truncated samples of data using maximum likelihood
arguments.  These methods are based on ideas first put forth by
Lynden-Bell (1971) and then further developed by Efron \& Petrosian
(1992).  Treatment of the most general case of multiply-truncated data
is discussed in Efron \& Petrosian (1999).  These non-parametric
statistical techniques used a well defined truncation criterion (and
the assumption that the observed sample is the one most likely to {\em
be} observed) to estimate the correlation between (if any), and
underlying parent distribution of, the relevant observables.  We now
describe these methods in the context of our particular problem A
particularly lucid explanation of the techniques as applied to quasar
data is found in Maloney \& Petrosian (1999).  These techniques
applied to GRB spectral data as well as simulations showing how
accurately these methods work can be found in Lloyd \& Petrosian
(1999) and Lloyd et al. (2000).
  
\subsection{How to Apply These Methods to our Data}
Consider a set of observables $L_{i}$ and $Z_{i}$ where $i$ indexes
the particular burst and in our case $i$ runs from 1 to 220.  The
luminosity suffers from a lower limit set by the constant flux
threshhold in the FRR sample, given by equation (1): $L_{lim}(Z) =
L(f_{256}=1.5 \rm ph/cm^2/s)$.  
\footnote{Alternatively, we can say that the redshift
$z$ suffers from an upper limit, given by the inversion of equation 1.
Either way we do
the problem gives us equivalent answers.}  For each burst indexed by
$i$, we can define an {\em associated set} $J_{i}$ defined as:
\begin{equation} J_{i} \equiv \{j: L_{j} > L_{i}, L_{lim,j} < L_{i}\}
\end{equation} This creates a truncation parallel to the axes as shown
by the horizontal and vertical dotted lines in Figure 2. The
associated set is all objects $j$ that could have been observed given
$i$.  Now, for each burst $i$, we can define a rank of $z_{i}$ in the
eligible set: \begin{equation} R_{i}=\#\{j\in J_{i}:z_{j} \le z_{i}\}
\end{equation}
 
We expect $R_{i}$ to be uniformly distributed between 1 and $N_{i}$,
where $N_{i}$ is the number of points in $J_{i}$.  To estimate the
degree of correlation between $L$ and $Z=1+z$, we can construct a
version of Kendell's $\tau$ statistic.  Let $T_{i} \equiv
(R_{i}-E_{i})/V_{i}$, where $E_{i} = (N_{i}+1)/2$, and $V_{i} =
(N_{i}^{2}-1)/12$.  Then, \begin{equation} \tau =
\frac{\Sigma_{i}(R_{i}-E_{i})}{\sqrt{\Sigma_{i}V_{i}}} \end{equation}
This $\tau$ statistic is normally distributed about a mean of $0$ with
a standard deviation of $1$ (Efron \& Petrosian, 1992).  Hence, a
$\tau$ of $-4$ implies a $4\sigma$ anti-correlation between the
variables at hand.  This is similiar to the usual Kendell's $\tau$
statistic defined as: 
\begin{equation} \tau_{K} =
\frac{pos-neg}{\sqrt{pos+neg+ytie}\sqrt{pos+neg+xtie}}
\left( \frac{4N+10}{9N(N-1)} \right)^{-1}
\end{equation} 
where $pos$ denotes the number of positive comparisons
(the ``position'' - meaning whether its greater than or less than - of
$x_{i}$ relative to $x_{j}$ is the same as the ``position'' of $y_{i}$
relative to $y_{j}$), $neg$ denotes the number negative comparisons
(the relative positions of the $x's$ and $y's$ are different), $tie$
denotes $x_{i}=x_{j}$ ($xtie$) or $y_{i}=y_{j}$ ($ytie$), and $N$ is
the number of data points.  This is also normally distributed about
$0$ with a standard deviation of $1$.  The difference between $\tau$
and the usual Kendell's $\tau_{K}$ is that, in the former case, points
are compared only if they are within eachother's truncation limits (in
other words, only points within associated sets are compared).  For
example, if $y_{j}$ fell below the lower limit of $y_{i}$, these two
points would not be compared.  Once the existence of the correlation
is established, it is easy to parameterize it in some quantitative
way.  One needs only to transform one of the variables, say $L$, as $L
\rightarrow L^{'}=L/\lambda(Z)$, where $\lambda(Z)$ can be written as
$\lambda(Z)=(1+z)^{\alpha}$, and then vary $\alpha$ until $\tau=0$.
 
It is also simple to estimate the underlying parent distributions of
the variables at hand once the correlation is known.  This method
relies on the assumption of independence of variables, so we must
apply this method to the uncorrelated variables $L'$ and $Z=1+z$.  For
uncorrelated variables $L'$ and $Z$, we again rely on finding the
associated set for each variable and the number of points in that set
$N_{i}^{'}$.  Then, the cumulative distribution $\Phi$ of $L'$, which
is the number of bursts with $L'$ greater than $L_{i}^{'}$, is given
by the simple formula (Lynden-Bell, 1971, Efron \& Petrosian, 1992,
Petrosian, 1993): \begin{equation} {\rm ln}\Phi(L_{i})=\sum_{j<i}{\rm
ln}(1+\frac{1}{N_{j}}) \end{equation} For each point indexed by $j$, a
truncation parallel to the axes is made and a weight $1/N_{j}$ based
on the number of points in the associated set is assigned to that data
point.
  
Simulations demonstrating how well these techniques work at
determining underlying correlations and the distributions given a well
defined truncation are in published in the Appendices of Lloyd and
Petrosian (1999) and Lloyd et al. (2000).  There, we simulate
correlated parent distributions, invoke a truncation representing some
observational selection effect, and define an observed sample by those
points that survived truncation.  We then 1) recover the correlation
present in the parent sample even when the truncation produces an
``observed'' correlation of the opposite sign, and 2) fully recover
the distributions of the underlying parent samples.

\section{Results}
We apply the statistical techniques described above to the 220 FRR
bursts shown in Figure 2.  We find that there is a significant
($>10\sigma$) correlation between the underlying parent distributions
of luminosity and redshift in this sample, when accounting for the
truncation. In other words, the null hypothesis that
luminosity is independent of redshift is rejected at the $10\sigma$
level.  The functional form of the correlation can be conveniently
parameterized by $\lambda(Z) \propto (1+z)^{1.4 \pm 0.2}$, in the
sense that $L^{'} = L/\lambda(Z)$ is uncorrelated with redshift (the
error bars here represent the statistical error for this
data set; see \S 4.1 for a discussion of the error due to the scatter
in the L-V relation).  The
function $\lambda(Z)$ therefore represents a sort of average
luminosity as a function of redshift\footnote{Although it may eventually be
useful to try other functions of luminosity evolution other than a
simple power law, such detailed analyses are not necessary
for the purposes of this paper and should wait until the
luminosity-variability correlation is more quantitatively sound. For
our purposes, the most important result is that the GRB luminosity
{\em does} evolve with redshift - we address the meaning and
robustness of this result in the text.}.  This is the first time luminosity
evolution has been demonstrated. To the best of our knowledge, all
results published previously regarding the luminosity and redshift
distributions of GRBs, which have implicitly assumed no
luminosity evolution in their analyses (e.g. FRR, Schmidt 2001, and
others).  We again
point out that our luminosity is actually modulo a burst beaming
factor $d\Omega$, so that this correlation may be an indication of
evolution of {\em either} the actual luminosity or the burst jet
angle. We address this issue in the discussion section below.
\begin{figure}[t]
\centerline{\epsfig{file=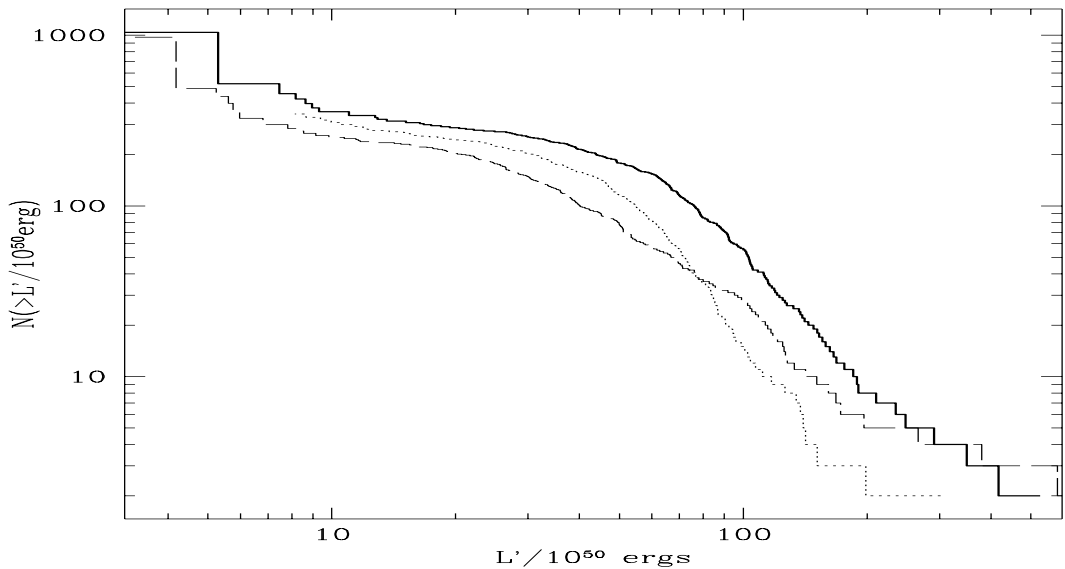,width=0.9\textwidth,height=0.49\textwidth,angle=0}}
\caption{Cumulative $L^{'}$ distribution $N(>L')$ as a function of $L^{'}$: for
220 bursts in the FRR sample, accounting for the flux threshold selection limit.
As described in \S 3 and 4, $L^{'}$ is the luminosity with the redshift dependence
$(1+z)^{1.4}$ removed: $L'=L/(1+z)^{1.4}$. The heavy solid line is for the best
fit $L-V$ relation, while the dotted and dashed lines are for the lower and upper
bounds to the relation, respectively (see \S 4.1 in text). }
\end{figure}
 
Since we have an estimate of the correlation present between $L$ and
$Z$, we can independently compute the distributions of the
uncorrelated variables $L^{'}$ and $Z$.  These distributions are
presented in various formats in Figures 3 through 7. Figure 3 shows
the cumulative distribution of $L^{'} = L/\lambda(Z) =L/(1+z)^{1.4}$ for
the best fit $L-V$ relation (solid line) as well as for the lower
(dotted line) and upper (dashed line) bounds to this relation (see \S 4.1). 
\begin{figure}[t]
\centerline{\epsfig{file=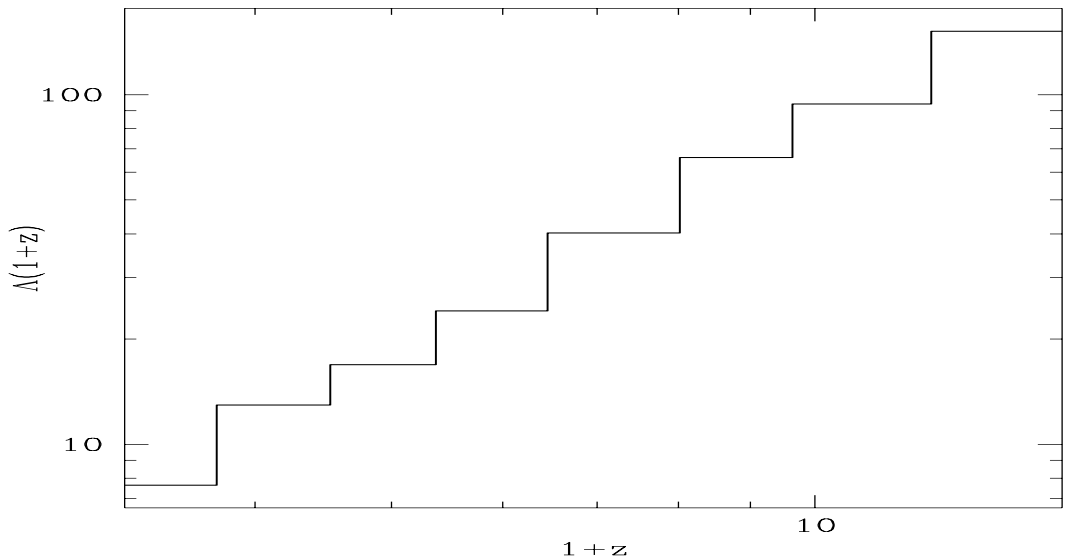,width=0.9\textwidth,height=0.48\textwidth,angle=0}}
\caption{ Total emitted luminosity as a function of redshift for the 220 GRBs
in the FRR sample, using luminosities and redshifts for the best fit L-V relation.}
\end{figure}
 We have tried several fits to this cumulative luminosity function (for
the solid line) with the following results: a single power-law fit to
the data yields $N(>L^{'}) \propto L^{'-1.28 \pm 0.02}$.  Fitting a
double power law to this curve we find indices $-0.51$ and $-2.33$
above and below respectively a break of $L^{'} = 5.9\times 10^{51}$
ergs.  These results yield a {\em differential} luminosity function
$dN/dL^{'}$ with indices $\sim -1.5$ in the shallower parts (low
$L^{'}$) to $\sim -3$ in the steepest parts (higher $L^{'}$).  This is
similar to the luminosity function found by FRR, who find a
differential luminosity distribution (without accounting for
luminosity evolution) of $dN/dL \propto L^{-2.3}$ for this sample.
However, because $L^{'}$ is not the actual luminosity of the burst,
but rather the luminosity with the redshift evolution removed, the
interpretation and comparisons of the luminosity function are not
straightforward.  Another, perhaps more useful, quantity to examine is
the total luminosity emitted at a given redshift.  The total
luminosity emitted at a given redshift is defined as ${\Lambda}(Z) =
\int_{0}^{\infty} L\Psi(L,Z)dL =
\rho(Z)\int_{0}^{\infty}\phi(L/\lambda(Z))L (dL/\lambda(Z)) = \rho(Z)
\lambda(Z) \int_{0}^{\infty}\phi(L')L' dL'$.  So that ${\Lambda}(Z)
\propto \rho(Z)\lambda(Z)$. As seen in Figure 4, this function
strongly increases with redshift (with a slope of about $2$) with no
apparent turnover out to a redshift of at least 10.

\begin{figure}[t]
\centerline{\epsfig{file=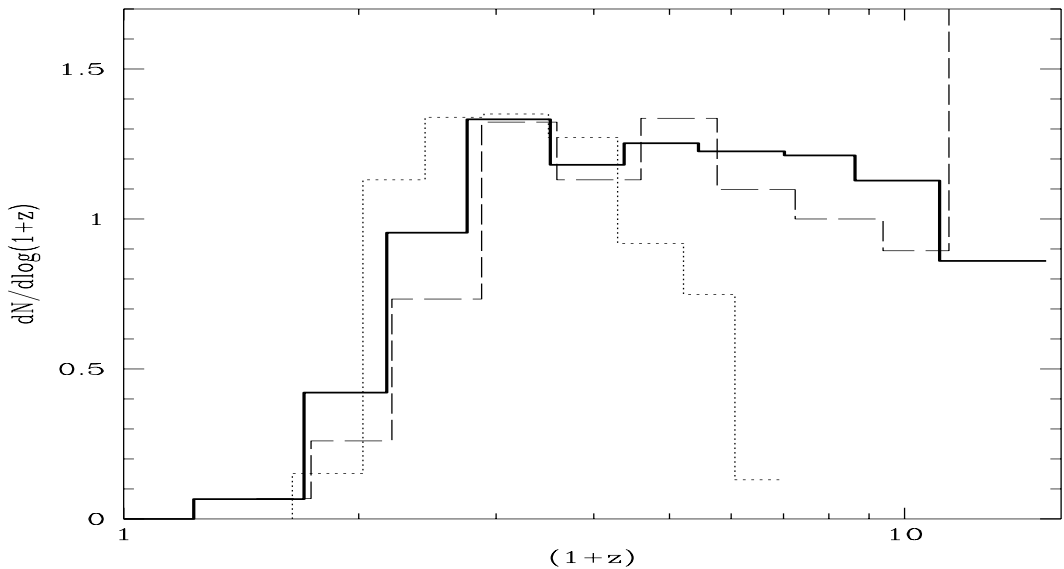,width=0.9\textwidth,height=0.49\textwidth,angle=0}}
\caption{Differential distribution of gamma-ray bursts in the FRR sample as a function
of redshift. The heavy solid line is for the best
fit $L-V$ relation, while the dotted and dashed lines are for the lower and upper
bounds to the relation, respectively (see \S 4.1 in text).}
\end{figure} 
Figure 5 shows the cumulative distribution of redshift, $N(<Z)$;
the solid line is for redshifts from 
the best fit value of FRR's $L-V$ relation, while the dotted
and dashed lines are the distributions for the lower and upper bounds
to the L-V relation, respectively.  
\begin{figure}[t]
\centerline{\epsfig{file=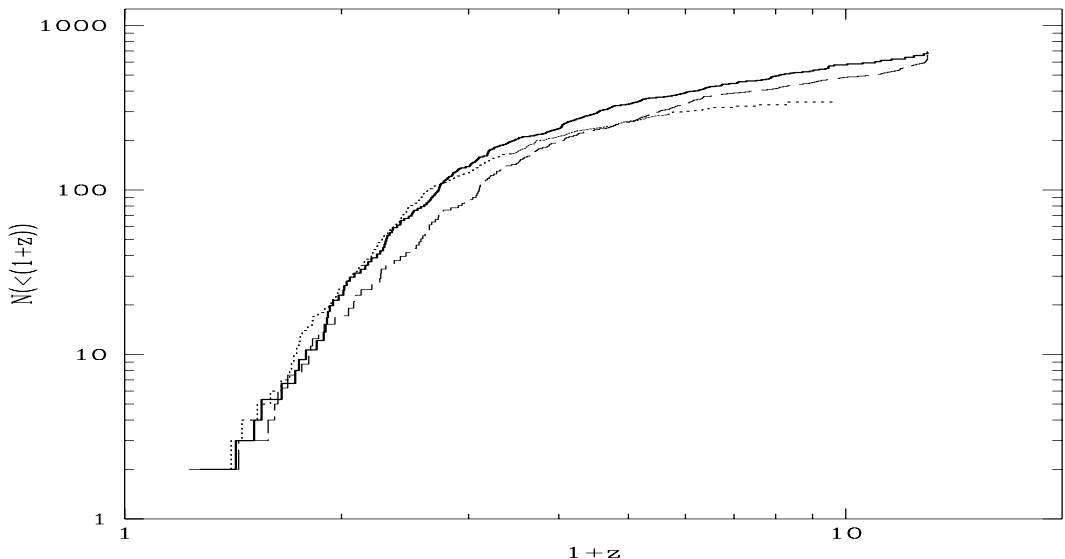,width=0.9\textwidth,height=0.49\textwidth,angle=0}}
\caption{ Cumulative distribtion $N(<1+z)$ as a function of redshift for the 220 bursts in
the FRR sample, accounting for the truncation threshold in the L-z plane.  The solid line represents
the cumulative distribution for the best fit value of the L-V relation.  The dotted
and dashed lines are the distributions for the lower and upper bounds to the L-V
relation, respectively (see \S 4.1 in text).  }
\end{figure}
It is evident that the
distribution of GRBs increases with increasing $z$.  In Figure 6, we
show the differential distribution $dN/dZ$ of GRBs as a function of
redshift. 
From this differential number distribution, we can derive a
co-moving rate density of gamma-ray bursts through the relation:
\begin{equation} \rho(Z) = \frac{dN}{dZ}(1+z)(\frac{dV}{dZ})^{-1},
\end{equation} where $V$ is volume, and \begin{equation} dV/dZ =
4\pi(c/H_{o})^{3}(\int_{1}^{1+z}\frac{d(1+z)}{\sqrt{\Omega_{\Lambda} +
\Omega_{m}(1+z)^{3}}} )^{2} \frac{1}{\sqrt{\Omega_{\Lambda} +
\Omega_{m}(1+z)^{3}}}.  
\end{equation} 
The additional factor of
$(1+z)$ in equation (7) comes from the fact that we are measuring a
{\em rate} and so must account for cosmological time dilation.  We
have plotted the density distribution in Figure 7 for
$\Omega_{\Lambda}=0.7$, $\Omega_{m}=0.3$, and $H_{o}=65 km/s/Mpc$,
where we have arbitrarily normalized the curve so that $\rho(Z)=1$ at
$Z=1$.  Note that the density sharply rises as $\sim (1+z)^{3}$ to a
redshift of about 2,  and then rises at a slower rate  
after that ($\sim (1+z)^{1}$) to at least a redshift of 10. In \S
5.2, we will use the GRB redshift distribution to place limits on the
global star formation rate at high redshifts.
\begin{figure}[t]
\centerline{\epsfig{file=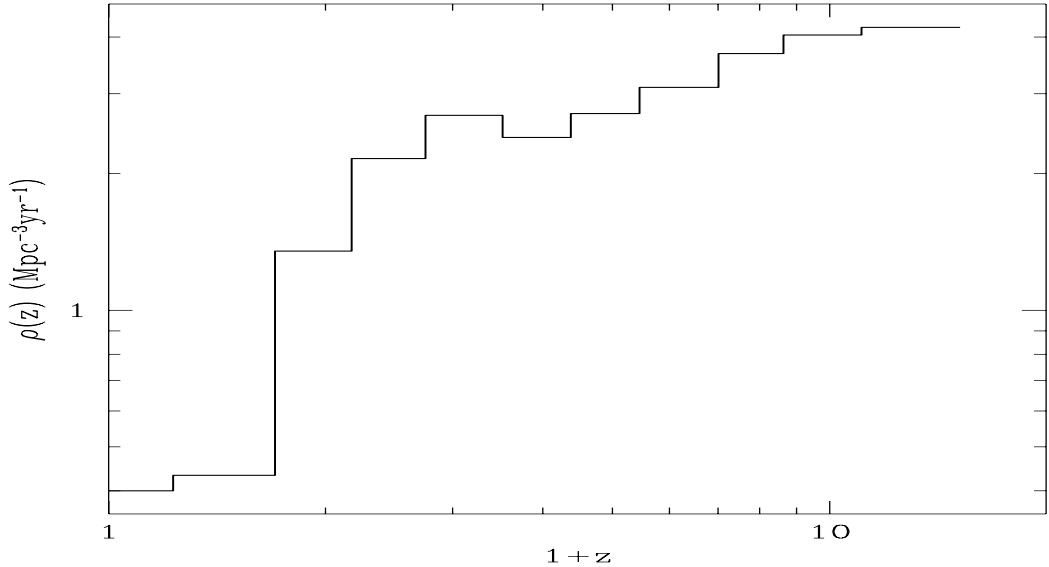,width=0.9\textwidth,height=0.49\textwidth,angle=0}}
\caption{The co-moving GRB rate density  $\rho(1+z)$ for our sample.
We have arbitrarily normalized the curve to $\sim 1$ at $(1+z)=1$.}
\end{figure}

\subsection{Robustness of the Results}
Because there is a significant amount of scatter in the $L-V$
relationship, it is important to discuss how our results change as a
function of the power law index in the $L-V$ correlation.  First, we have
repeated our analysis using tables of redshifts and luminosities from
the two bounding relationships in FRR: $L \propto V^{1.03}$ and $L
\propto V^{2.05}$.  In both cases we find that the null hypothesis
of no luminosity evolution is rejected with very high
significance - $6.5 \sigma$ and $6 \sigma$ for the lower and
upper bounds, respectively. 
 We can parameterize the evolution in
these cases as ${L} \sim (1+z)^{1.3 \pm
0.2}$ for the lower bound and ${L} \sim (1+z)^{1.9 \pm 0.2}$ for the
upper bound.  In addition, we have attempted to account for
the dispersion in the L-V relationship in the following way: for
a given observed variability $V$, we draw a value for
the luminosity from a uniform distribution 
covering the total possible range of luminosities allowed 
by the bounding power laws of the L-V relation.
  We then compute the corresponding redshift (given
our cosmological model described above).  We then
test whether the null hypothesis of no correlation between
luminosity and redshift holds.  Once again, we find that it
is rejected with high significance ($\sim 6 \sigma$), and that
the evolution can be parameterized as $L \sim (1+z)^{1.7 \pm 0.4}$.
We emphasize once more that we have no a priori assumptions about the
parametric form of the underlying data nor the relationship
between luminosity and redshift. 

Figure 3 shows the cumulative luminosity distribution $N(>L')$, for
luminosities dervied from the upper limit (dashed line) and
lower limit (dotted line) to the $L-V$ relation.
 The luminosity functions for the ``bounding'' data are
similiar to one another, breaking at around $5 \times 10^{51} \rm ergs/s$,
as with the best fit value; however, the lower bound to the $L-V$
relation gives a more narrow luminosity function, while the upper
bound broadens it relative to the best fit relation. Figures 5 and 6
show the cumulative and differential distributions, $N(<Z)$ and
$dN/d{\rm log}Z$ respectively, for the bounding L-V relations. It is clear from
Figures 5 and 6 that the lower bound (dotted line) to the $L-V$
relation places the large majority of bursts at relatively low
redshifts (below $z \sim 4$), while the upper bound (dashed line) to
the $L-V$ relation causes a disproportionate number of bursts to be
located at $z \ga 10$.  In the remainder of this paper, we will
utilize the nominal values of the
luminosities and redshifts as derived in FRR.
 We also point out that the
dependence on the choice of cosmological model is not a significant
source of uncertainty (compared to the uncertainty from the scatter in
the L-V relationship).   

\section{Discussion}

\subsection{On the Presence of Luminosity (or Jet Angle) Evolution}
Under the assumption that the luminosity-variability correlation
indeed produces valid redshifts for GRBs, the data show the presence
of significant luminosity evolution (a $> 10\sigma$ correlation
between $L$ and $1+z$) in which the average GRB luminosity scales as
$(1+z)^{1.4}$.  Even when the slope of the underlying
luminosity-variability relation changes (within the maximum bounds
allowed by the 8 GRBs used to derive
this relation), the correlation between GRB luminosity and
redshift remains.  It is important to address the question of the
origin or meaning of this evolution.  Because our variable $L$ above
is really $L/d\Omega$, where $d\Omega$ is the jet opening solid angle
of the GRB, the correlation of $L$ with redshift can result from the
evolution of either $L$, or $\Omega$ or both.  We discuss each in turn
below.
  
\subsubsection{Luminosity Evolution}  

If the correlation between $L/d\Omega$ and redshift is due to the
evolution of the amount of energy per unit time emitted by the GRB
progenitor (and not evolution of the jet opening angle), then this
suggests that bursts were significantly brighter in the past.  
Presumably, this evolution is the result of a variation in 
the progenitor of the GRB.  If the progenitors of GRBs are 
indeed massive stars, then somehow the distribution or structure 
of these massive stars must be different at high redshifts.

Currently, both observational and theoretical arguments (e.g. Larson,
1998, and references therein; see also Malhotra \& Rhoads, 2002)
 suggest that the stellar initial mass function
(IMF) was ``top heavy'' at high redshift - that is, the mass scale of
the IMF was higher in the earlier stages of the universe. A time
varying IMF with a higher mass scale at higher $z$ is a very realistic
possibility based on the following arguments from Larson, 1998: The
mass scale in the IMF is likely to depend strongly on the temperature
in star forming clouds.  In the early universe, this temperature was
probably higher for several reasons - the cosmic background
temperature was higher, the metallicity was lower which implies lower
cooling rates and therefore higher temperatures on average, heating
rates were probably higher in the past because the star formation rate
per unit volume was higher leading to more intense radiation fields at
high redshifts (for further details, see Larson, 1998).  The higher
temperatures imply that a larger mass scale (e.g. Jeans mass) is
required to collapse protostellar material and form a star. Because of
these and other effects, Larson suggests that the mass scale of the
IMF could have been as much as an order of magnitude or more higher at
redshifts larger than 5.  

In addition, since mass loss from stellar winds seems to depend on stellar
metallicities (e.g. Meynet et al. 1994; Langer \& Henkel 1995; Ramirez-Ruiz
et al., 2000), these
massive stars are less likely to lose much mass before collapse.  
At low redshift (roughly solar metallicities), even if these massive 
stars formed, they would lose most of their mass due to winds before 
collapse.  So at higher redshift, the progenitors of GRBs are not only 
likely to form with higher average masses, but will also retain this 
mass until collapse.

One might expect that this increase in mass could lead to an overall
larger energy budget and hence more luminous bursts at earlier times.
For example, a fairly generic scenario for the engine which drives
GRBs invokes the rapid accretion of a disk around a stellar-massed
black hole (Popham, Woosley, \& Fryer 1999).  In such a scenario, the
energy in the burst increases dramatically with amount of mass in, and
accretion rate of, the disk (Popham et al. 1999).  More massive progenitor
stars tend to produce higher accretion rates and hence, stronger GRBs.
In general, this is true whether the engine producing the burst is 
driven by neutrino annihilation or magnetic fields.  

The simple argument of ``more massive progenitor = more luminous
GRB'', however, ignores much of the subtle physics required to produce
a GRB from some cataclysmic event (such as a collapse to a black hole
or some sort of compact object merger).  For collapsar models, the
{\em most massive} ($300 M_{\odot}$) stellar progenitors may only
produce weak or no relativistic outflows (Fryer, Woosley, \& Heger
1999).  Although the accretion rate on the black hole is much higher
for these stars, the angular momentum is too low to produce a stable
disk until most of the star has already accreted onto the black hole.
The characteristics of a GRB outflow - particularly in these collapsar
models - also depend on the density of the surrounding stellar envelope
which the outflow must "punch" through, the accretion rate (which
depends on many factors including the spin of the BH), the presence of
magnetic fields, etc. Of course, the luminosity of the GRB outflow
also relies on the mechanism and efficiency of energy extraction from
the black hole, which (for example in a Blandford-Znajek mechanism;
Blandford \& Znajek, 1977; see Popham et al. 1999 for a review) can
depend critically on the angular momentum as well as the mass of the
black hole.

The relationship between progenitor mass and GRB energy output is
therefore not straightforward for all progenitor models.  
{\em However, there are some situations in which this relationship holds
very well}.  If the specific angular momentum in the core of massive
stars did not vary with stellar mass, then it is likely that more
massive stars would produce more energetic bursts (MacFadyen \&
Woosley, 1999).  For some black hole accretion disk models (e.g. He
mergers), Zhang \& Fryer (2001) have shown that the GRB energy does
increase with increasing stellar mass.  Therefore, depending on the
model, it is possible that progenitor mass evolution (due to a
top-heavy IMF) or core mass evolution (due to metallicity effects) may
be responsible for the observed luminosity-redshift correlation.  But
this is not the only possibility as we discuss below.
  
  
\subsubsection{Jet Opening Angle Evolution}
It was recently suggested (Frail et al., 2000; see also Panaitescu \&
Kumar, 2001) that the GRB luminosity is in fact constant, and it is
the jet opening angle that causes the dispersion in the apparent or
isotropic equivalent GRB luminosity.  This analysis relies on the fact
that the break observed in many GRB afterglow light curves is due to a
geometrical jet effect (Rhoads, 1997) and not, for example, a
transition from a relatvistic to non-relativistic phase (Huang, Dai \&
Lu, 2000) or environmental effects such as a sharp density gradient
(Chevalier \& Li, 2000; Ramirez-Ruiz et al., 2000).
  If the Frail et al. result is correct, then
our results above imply the existence not of luminosity evolution, but
dependence of the {\em jet angle} as a function of redshift.  In this
case, we trade $L$ for $1/\Omega$ so that the jet solid angle $\Omega$
is proportional to $(1+z)^{-1.4}$; in other words, the average opening
angle is expected to be smaller in the past\footnote{Note that all of the
distributions for $L$ translate into distributions for $1/\Omega$ via
a factor $\sim 10^{50}$ ergs, the Frail et al. standard luminosity.}.
  
Physically, the evolution of the jet opening angle is a matter of
speculation.  The angle into which the baryonic/leptonic outflow is
beamed is probably strongly dependent on, among a number of factors, the rotational
velocity of the progenitor. For example, in the collapsar model of
MacFadyen \& Woosley (1999), a GRB is produced when a very massive
star with angular momentum in a specified range (enough to prevent
spherical collapse, but not too much to prevent collapse on a
reasonable timescale) collapses to a black hole.  Because the star
is rapidly rotating, accretion is suspended along and around the
equitorial plane. 
 Thus, only matter along the rotation axis can fall into the black
hole and evacuate a region to allow for reversal of flow - that is,
outflow from the progenitor, which will produce the GRB jet.  For
higher rotational velocities, this evacuated
region will be more collimated and thus have a smaller jet opening angle. Therefore,
evolution of the progenitor rotational velocity (in which stars
rotated more rapidly at higher redshifts) could at least qualitatively
explain our results\footnote{We point
out that this model does not include MHD, which is also
likely to play an extremely important role in the collimation of the
jet.}.  However, it is not at all clear how one might
produce this type of evolution.   
  
\subsubsection{Comparison with Other Luminosity Functions.}
Not suprisingly, the luminosity functions of many astrophysical
objects (e.g. galaxies, quasars, etc.) show signs of evolution with
redshift.  It is often a difficult task to quantify this evolution -
observationally and theoretically - due to either numerous selection
effects (e.g.  dust, malmquist biases, sensitivity, etc.) which tend
to plague the observations, or the presence of simplifying assumptions
and lack of numerical resolution in the theoretical simulations.
However, we can make some quantitative statements about the luminosity
evolution of various astrophysical objects.  For instance, it is
fairly well established (Marzke et al. 1994) that the local luminosity
function varies for early and late type galaxies; when parameterizing
the luminosity function by the standard Schechter function, the power
law index evolves from $\alpha=-0.8$ for early type galaxies to
$\alpha=-1.8$ for late type galaxies (implying more luminous galaxies
at earlier times).  Fried et al. (2001) find that the luminosity
evolution depends significantly on the galaxy type, and present a list
of luminosity functions and their evolution for many galaxy types (see
Table II in their paper).  In the case of starburst galaxies, for
example, the highest redshift galaxies are intrinsically brighter than
the lowest redshift, but fainter than intermediate redshift.
Numerical simulations of Nagamine et al. (2001) find that the
characteristic luminosity of galaxies increases by 0.8 mag from $z=0$
to $2$ and then declines towards higher redshift, while the B-band
{\em luminosity density} continues to increase from $z=0$ to $5$
(although only slowly after $z \sim 3$), for a $\Lambda$ CDM universe.
The quasar luminosity function also appears to evolve with redshift.
The luminosity evolution is generally accepted as $L \sim (1+z)^{3}$
for $z<1.5$ and constant after that up to redshifts of 3 (Boyle, 1993;
Hewett et al. 1993), although alternative models suggest $L \sim
(1+z)^{1.5}$ out to a redshift of 3 (Hewett et al., 1993).

Comparison of these luminosity functions with the GRB luminosity
evolution is not straightforward because - as mentioned above - of the
various selection effects that play a role in determining galaxy and
quasar luminosity functions, not to mention the fact that it is
unclear how one would physically connect the GRB to these types of
luminosity functions.  It would be perhaps most useful to compare our
results for GRB luminosity evolution with that of different types of
supernovae, because of the eminent link between GRBs and massive
stars.  Unfortunately, this will have to await the next generation of
supernovae observations (such as SNAP), while the current data is
insufficient to say anything about the evolution of supernovae
luminosity to high redshifts. It should be emphasized again, however,
that almost all astrophysical objects for which sufficient data exists
have shown evidence of luminosity evolution, and its theoretical
origin as well as observational consequences should be seriously
investigated for GRBs.

\subsection{The GRB Density Distribution and Estimating the Star Formation
Rate at High Redshift} 
There has recently been a large volume of both observational and
theoretical work suggesting GRBs are associated with massive stars.
The observational evidence is based on colors and other properties of
GRB host galaxies suggesting active star formation (Sokolov et
al.,2001), burst locations in their host galaxies (Bloom et al.,
2000), possible evidence of supernovae in the light curves of some
afterglows (Bloom et al., 1999, Reichart, 2000), and the and direct
iron line observations in the X-ray afterglows of a few GRBs (Piro et
al., 2000 Antonelli et al., 2000)\footnote{We point out, however, that some
studies of extinction in the light curves of GRBs (e.g. GRB000926)
show a different extinction than what is expected for young star
forming regions (Price, et al., 2001).  In addition, Tsvetkov et
al. (2001) find that there is a small probability ($4\%$) that GRBs
and star forming sites belong to the same region.}.   Theoretically, massive star progenitors
have proved to be viable models particularly for the longer duration
bursts (Woosley, 2000, MacFadyen \& Woosley, 1999). If GRBs are indeed
the final episode in a single massive star's life, then the rate
density of GRBs will be strongly correlated with the overall star
formation rate. Other progenitors such as binary mergers will also
show evidence of such a correlation, although the correlation is not
as straightforward (see \S 5.2.1 below; see also Fryer, Woosley, \&
Hartmann, 1999). Other arguments
linking GRBs with massive stars can be found in Totani (1997), Blain
\& Natarajan (2000), Lamb \& Reichart (2000), and Ramirez-Ruiz,
Trentham, and Blain (2001).
 
Therefore, GRBs are in principle ideal tools for constraining
the star formation rate, which has 
been one of the most
important problems in extra-galactic astrophysics for decades.
  One of the advantages of
using GRBs to trace the SFR is that the gamma-rays and X-rays from the
burst travel from the source to our telescope relatively unhindered
and so are not subject to the types of selection effects present in
the UV and Far-IR methods of determining the SFR (see Schaerer, 1999,
for a review). Indeed using the luminosity-variability correlation to
determine redshifts relies on the gamma-ray data, which avoid
selection effects present in lower energy bands.  Furthermore, because
GRBs may exist out to redshifts $z \ge 10$ if they can be associated
with the earliest population of massive stars, they can probe the SFR
to higher redshifts than any other method used previously (see also
Lamb \& Reichart, 2000).
 
In Figure 7, we have plotted the co-moving rate density distribution
as a function of redshift for an $\Omega_{\Lambda}=0.7$,
$\Omega_{M}=0.3$ universe.  First, we note that there is an increase
in the rate density $\sim (1+z)^{3}$ out to a redshift of $(1+z) \sim
3$.  Although there are relatively few bins or points here, this is
similar to independent determinations of the star formation rate
to this redshift using various methods (as well as a ``meta-analysis'' of the
SFR combining the various methods; see Hogg, 2001 and references
therein).  Beyond a redshift of $(1+z) = 3$, the GRB rate density
flattens, although continues to rise as $\sim (1+z)^{1}$ until at least
a redshift of 10, beyond which our sample is truncated.  
 The very large portion of the co-moving rate density at high
redshift may be evidence of a sample of population III stars.  In the
following section, we address the issues involved with associating the
GRB rate density to that of the star formation rate, and then make
estimates of the SFR based on different progenitor models for GRBs.
       
\subsubsection{Estimating the Star Formation Rate from the GRB Rate}
As mentioned above, most gamma-ray burst models are connected to the
formation of massive stars and for some mechanisms, it is likely that
the gamma-ray burst rate traces the SFR in the universe (Fryer,
Woosley, \& Hartmann 1999).  However, as we probe increasingly higher
redshifts, we find that metallicity dependent effects (mass loss from
winds, slope of the initial mass function) can change the number of
GRBs produced for a given amount of star formation (especially for the
collapsar model which arise from only the most massive stars).  In
addition, the delay between stellar formation and merger for compact
merger scenarios (double neutron star, black hole/neutron star)
produces a detectable shift at high redshift.

To extract a star formation rate at high redshift from the gamma-ray
burst distribution, we must not only know which progenitor(s) produce
gamma-ray bursts, but how the evolution of these progenitors varies
with increasing redshift.  Although deriving an exact star formation
rate is impossible at this time because of uncertainties (such as
rotation rates of massive stars or neutron star kicks), with a few
assumptions we can place some interesting limits on the star formation
history using gamma-ray bursts.  At one extreme, we take the collapsar
model whose rate of formation for a given star formation rate
increases dramatically with redshift, both because the amount of wind
mass-loss decreases and because the initial mass function flattens at
high redshift.  At the other extreme, we consider the double neutron
star and black hole-neutron star mergers which, because of the delay
in their merging, produce fewer bursts for a given star formation rate
at high redshift.
 
For both model scenarios (collapsers and binary mergers), we assume 
that mass loss from winds and the effects of the initial mass function
are the only redshift dependencies on stellar evolution.  For the
collapsar model in particular, we are assuming that the rotation rate
of a star at collapse does {\it not} depend on redshift and that the
structure of the core depends only on mass loss metallicity effects.
Stellar evolution models confirm the latter assumption, but too little
is understood about rotation to judge our first assumption at this
time.  For binary merger scenarios, we assume that the kick imparted 
onto neutron stars and black holes at birth does not vary with redshift.

For mass-loss, we assume that the mass loss rate ($\dot{M}_{\rm Winds}$) 
depends on the metallicity:
\begin{equation}
\dot{M}_{\rm Winds} \propto z_{\rm metal}^{0.5},
\label{eq:wind}
\end{equation}  
where $z_{\rm metal}$ is the metallicity (in units of solar
metallicity).  For the metallicity as a function of redshift ($z$), we
are guided by the distribution from Pei, et al. (1999):
\[ z_{\rm metal}= \left\{ \begin{array}{ll} 
                        10^{-0.5z} & {\rm for}\; z<3.2 \\
                        10^{-0.8-0.25z} & {\rm for}\; 3.2<z<5.0 \\
                        10^{-0.2-0.4z} & {\rm for}\; z>5.0 
\end{array} \right. \] 
Given the metallicity as a function of redshift, we can determine the
mass loss rate from equation 9. 
\begin{figure}[t]
\centerline{\epsfig{file=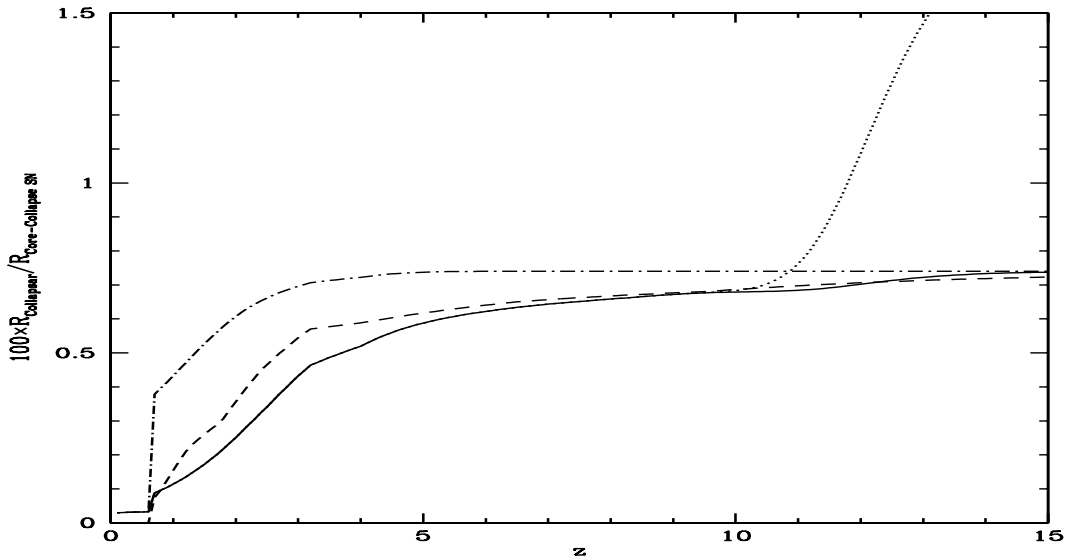,width=0.9\textwidth,height=0.49\textwidth,angle=0}}
\caption{Collapsar rate as a function of redshift.  The solid line assumes
that the IMF does not change with redshift, but that mass loss from
winds does change with metallicity: $x=0$ above half solar metallicity
and $x=0.5$ at high redshift when the metallicity is below 1/2 solar
(see \S 5.1 for details).  For the solid line, the metallicity at a
given redshift is given as a gaussian with a $1-\sigma$ deviation set
to 0.5 in the log of the metallicity.  The dashed line gives the rate
as a function of redshift without any spread in metallicity.  The main
effect of the spreading occurs at low redshift.  The dotted line
is identical to the solid line, but the IMF is flattened for population
III stars.  The dot-dashed line is the rate if we assume the mass loss
changes more abruptly below 1/2 solar metallicities ($x=2$).}
\end{figure}
 Unfortunately, the
crucial parameter in calculating the formation rate of gamma-ray burst
progenitors is the total mass lost, not the mass loss rate.  Because
mass loss alters the lifetime and luminosity of the star, the total
mass loss, $\Delta M_{\rm Total}$ does not scale linearly with the mass loss rate.  Depending
on the stellar model, $\Delta M_{\rm Total} \propto \dot{M}_{\rm
Winds}^x \propto z_{\rm metal}/2$ where $x$ lies somewhere between
0.2-2.0 (e.g. Meynet et al. 1994; Langer \& Henkel 1995).  From Meynet
et al. (1994), we see that for a $40M_{\odot}$ star, the total mass
lost to winds does not decrease significantly ($x \approx 0$) until
the metallicity drops below one half of solar, at which point it drops
dramatically ($x \approx 0.5$ at $z_{\rm metal}=0.2$ and $x \gtrsim
1.0$ at $z_{\rm metal}<0.1$).  For our calcuations, we assume that
$x=0$ for metallicities above one half solar.  At lower metallicities,
we use a range of values for $x$ to determine the dependence of the
collapsar rate on this uncertain quantity (see Figure 8).  The
dependence of mass loss on metallicity is the dominant uncertainty in
our calculations.  In general, at any given redshift, there will be a
distribution in metallicities which will tend to dilute the effects of
metallicity variation as a function of redshift.  However, as we see
in Figure 8, this dilution only significantly alters the formation
rate at low redshifts ($z<1$) and is not crucial for our study of the
high redshift star formation rate\footnote{However, if the mass loss
rate remains relatively flat down to metallicities of one tenth solar
or lower, metallicity effects can be important at high redshifts.}.

The last effect that we include is the flattening of the initial 
mass function at high redshifts.  It has long been thought that 
the first generation of stars (so-called population III stars) 
had an initial mass function skewed to more massive members 
(e.g. Silk 1983; Carr \& Rees 1984; see also \S 5.1.1).  To test the limits of this, 
we assume that population III stars dominate the stars formed 
above a redshift of 5 and that the initial mass function flattens 
from a steep IMF ($f(M)\propto M^{-2.7}$, where $M$ denotes mass) at low redshift to a very 
flat IMF ($f(M)\propto M^{-1.5}$) for redshifts above 5.  The 
effect of this IMF flattening on the collapsar rate is shown in Figure 8.  

Figure 9 shows the co-moving rate density derived from both sets of
progenitor models (using our results from
the ``best-fit'' L-V relation).  We have assumed
that that $x$, the metallicity dependence, is 0,0.5 for respective
metallicities above, below one-half solar.  We have also assumed that
there is a Gaussian spread in the metallicity (with a $1-\sigma$
deviation set to 0.5 in the log of the metallicity).  The solid and
dotted lines in Figure 9 correspond to with and without the effects
of a flattening IMF in the collapsar model.  This provides a
lower-limit for the slope of the SFR (with the best-fit data).  The
dashed and dot-dashed lines correspond, respectively, to the same
models (with and without IMF flattening) for the compact merger
models.  These merger models provide a rough upper limit for the SFR
derived from the rate of GRBs (He-merger and supranova GRBs will lie
somewhere between the collapsar and compact merger burst models).  For
reference, we have superposed on the plot the star formation rate
(SFR) from Blain (2001; dark solid line) which includes contributions
to the SFR from both the UV and the sub-mm wavelength range (the
latter contribution from the reprocessed UV emission absorbed and
re-emitted by dust in the IR).
\begin{figure}[t]
\centerline{\epsfig{file=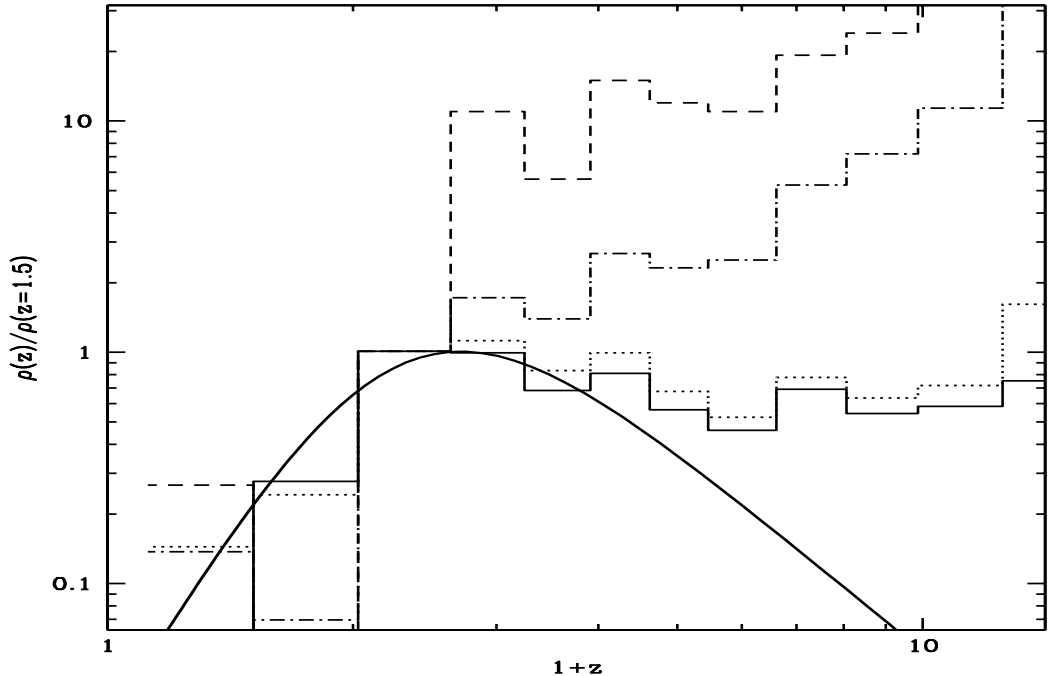,width=0.9\textwidth,height=0.6\textwidth,angle=0}}
\caption{Estimates of the star formation rate based on various progenitor models.
The solid and
dotted lines correspond to the collapsar model, with and without the effects
of a flattening IMF respectively.  This provides a
lower-limit for the slope of the SFR (with the best-fit data).  The
dashed and dot-dashed lines correspond, respectively, to the same
models (with and without IMF flattening) for the compact merger
models. We have
superposed on the plot the star formation
rate (SFR)
from Blain  (2001; dark solid line) which includes contributions to the
 SFR from both the UV and the
sub-mm wavelength range.}
\end{figure}

These calculations show the great potential that GRBs have for 
understanding the SFR at high redshifts.  Understanding the 
progenitors of GRBs allows us to extract firm lower limits 
on the star formation history at high redshifts.  Assuming the current 
best fit for the L-V relation, we find that the co-moving rate density 
does not decrease significantly with increasing redshift, and it 
may even increase.

These studies also provide evidence that compact mergers 
can not explain all GRBs.  To explain the number of high-redshift 
merger GRBs (which require a lot of quickly merging compact binaries), 
our models tended to overproduce the number of GRBs that occur 
at low redshifts.  If the rate of GRBs continues to increase
at high redshifts as our results suggest, and if 
there is not some redshift dependence on the compact object kick 
magnitude, we can rule out compact merger models as the sole 
source of gamma-ray bursts.

\section{Summary and Conclusions}
Presently, there are $\sim 20$ gamma-ray bursts with directly measured
redshifts. And although we have learned a remarkable amount from these
few bursts, we still know very little about the GRB
luminosity function and density distribution - two very important
pieces in the GRB puzzle.
Recently, it has been suggested (Norris et al., 2000; Fenimore \&
Ramirez-Ruiz, 2000; Reichart et al., 2000) that certain GRB observables
can be used as ``standard candles'' or luminosity indicators
from which to infer luminosities
and redshifts.  In particular, Fenimore \& Ramirez-Ruiz (2000; FRR) have
shown, using 8 bursts with measured redshifts,
 that there is a significant correlation between the luminosity
of a GRB and the variability of its light curve (the ``L-V relationship'').
  Using variability
measurements from the light curves of 220 BATSE bursts, they used the L-V
relation to compute a luminosity and (with knowledge of the observed flux and
assuming a cosmological model) redshift for each burst.

 In this paper, we have analyzed the FRR sample of 220 luminosities
and redshifts.  Because the FRR sample has a flux limit of 
$1.5 \rm ph/cm^{2}/s$, there is a strong truncation in the luminosity-
redshift plane, which makes it diffcult to directly determine
the underlying luminosity and redshift distributions of the parent
burst population.  We have used well developed non-parametric
statistical techniques to account for this selection effect.  Our
main results are as follows:

$\bullet$  {\bf We have shown that there exists a significant correlation between
the luminosity of a GRB and its redshift}, which
can be parameterized by $L \propto (1+z)^{1.4 \pm \sim 0.5}$.
This result is not very sensitive to the exact parameterization
of the luminosity-variability correlation - when taking the extreme values
for the power law indices of the 
the L-V relation or when drawing from a distribution of luminosities
(spanning the range allowed by the L-V relation),
we still find strong luminosity evolution,
although with slightly different redshift dependence. If the L-V relation
does indeed produce valid luminosities and redshifts for GRBs, then
the evolution of GRB luminosity with redshift
is a very robust result.

$\bullet$ We have estimated the GRB luminosity function {\em
with the redshift dependence removed}, $\phi(L') =
\phi(L/(1+z)^{1.4})$.  We find the luminosity function has
an average power law index of about $-1.3$, and an apparent break
at about $L' = 5 \times 10^{51} ergs$.  We also estimate the total luminosity
emitted as a function of redshift and find that this quantity increases
strongly with redshift (due not only to the large number of sources at high redshift, 
but also the strong correlation between luminosity and redshift).
 
$\bullet$ Accounting for the correlation between luminosity and redshift,
we have estimated the co-moving rate density, $\rho(1+z)$, of GRBs in this
sample.  We find that $\rho$ increases $\sim (1+z)^{3}$ until a redshift
$1+z \sim 3$, consistent with star formation density results to
this redshift (see Hogg et al., 2001).  {\bf Remarkably, the 
GRB rate density continues to increase as $\sim (1+z)^{1}$
 until a redshift of $(1+z) \sim 10$}, beyond which our sample truncates.
 
$\bullet$ From the GRB redshift distribution, we have used the population
synthesis codes of Fryer et al. (1999) to compute an estimate of the
star formation rate at high redshifts.  {\bf We find that the star formation
rate increases or remains constant to very high redshifts, no matter
what the progenitor model}.  We also find because our GRB rate
density increases to high redshifts, merger models - which tend to
overproduce GRBs at low redshifts - cannot be the sole
source for all GRBs.

In this paper, we have also speculated on the possible origin
of the luminosity evolution, and compared it to other known sources
of luminosity evolution.  The evolution we observe may be due to
either an evolving energy output of the GRB progenitor or a
changing jet opening angle with redshift.  We have discussed some
possible scenarios for each type of evolution; particularly, 
a top-heavy IMF may be responsible for producing more massive 
BH cores of collapsed stars and high redshift and this may lead
to more energetic bursts.  These speculations need to be examined
in further detail.

We emphasize again that our results rely on the existence of the L-V
relation and its ability to give correct luminosities and redshifts.
Future redshift measurements will help confirm and more quantitatively
establish this relationship.  However, a true test of our results will
have to await the Swift satellite, which is expected to get redshifts
to hundreds of bursts over a few years.  With a good handle on the
selection effects of the detector, multi-wavelength GRB observations
will not only provide us with a multitude of redshifts and
luminosities, but will help us constrain the GRB environment and
hopefully help elucidate the source of the luminosity evolution, and
its relationship to the GRB progenitor(s).

{\bf Acknowledgements:} We are very indebted to Ed Fenimore for
providing unpublished tables of redshifts and luminosities from the
FRR paper. We also are grateful to an anonymous referee whose
comments led to many clarifications in this paper.
 N.L-R. and C.L.F. would like to thank the Aspen Center for
Physics for their hospitality and for providing a good working
environment where some of this work was carried out.  We would also
like to thank Vahe' Petrosian, Don Lamb, Andrew MacFadyen, Shiho Kobayashi, 
 for useful discussions. N.L.-R. acknowledges support
from NASA grant NAG5-7144. The work of C.L.F. was funded 
by a Feynman Fellowship and an in-house ASCI grant at LANL.
E.R.-R. acknowledges support from CONACYT, SEP and the ORS foundation,
and would like to thank A. Blain, D. Lazzati, P. Natarajan \& N.
Trentham for  helpful conversations.

\end{document}